\documentclass{tcibook}
\usepackage{fancyhea}
\usepackage{work}
\usepackage{bm}       
\usepackage{graphicx}
\usepackage{hyperref}      

\begin{document}

\def\bibname{References}
\bibliographystyle{plain}

\raggedbottom

\pagenumbering{roman}

\parindent=0pt
\parskip=8pt
\setlength{\evensidemargin}{0pt}
\setlength{\oddsidemargin}{0pt}
\setlength{\marginparsep}{0.0in}
\setlength{\marginparwidth}{0.0in}
\marginparpush=0pt


\pagenumbering{arabic}

\renewcommand{\chapname}{chap:intro_}
\renewcommand{\chapterdir}{.}
\renewcommand{\arraystretch}{1.25}
\addtolength{\arraycolsep}{-3pt}

\chapter{Computing Frontier: Accelerator Science}
\label{chap:mag}
\begin{center}\begin{boldmath}
\begin{center}

\begin{large} {\bf Conveners: P. Spentzouris (FNAL), E. Cormier-Michel (Tech-X Corp.)\\
                   Observer: C. Joshi (UCLA)} \end{large}

J. Amundson (FNAL),
W. An (UCLA),
D.L. Bruhwiler (University of Colorado, Boulder),
J.R. Cary (Tech-X Corp.),
B. Cowan (Tech-X Corp.),
V.K. Decyk (UCLA),
E. Esarey (LBNL),
R.A. Fonseca (IST),
A. Friedman (LLNL),
C.G.R. Geddes (LBNL),
D.P. Grote (LLNL),
I. Kourbanis (FNAL),
W.P. Leemans (LBNL),
W. Lu (UCLA, Tsinghua University),
W.B. Mori (UCLA),
C. Ng (SLAC),
Ji Qiang (LBNL),
T. Roberts (Muons Inc),
R.D. Ryne (LBNL),
C.B. Schroeder (LBNL),
L.O. Silva (IST),
F.S. Tsung (UCLA),
J.-L. Vay (LBNL),
J. Vieira (IST)

\end{center}

\end{boldmath}\end{center}

\section{Introduction}
\label{sec:acc-sc-intro}
Particle accelerators are critical to scientific discovery both nationally and worldwide. The development and optimization of accelerators are essential for advancing our understanding of the fundamental properties of matter, energy, space and time. Modeling of accelerator components and simulation of beam dynamics are necessary for understanding and optimizing the performance of existing accelerators, for optimizing the design and cost effectiveness of future accelerators, and for discovering and developing new acceleration techniques and technologies.

The requirements for high-fidelity computer simulations of accelerator systems and accelerator components are driven by the need to develop and optimize new accelerator concepts and design machines based on these concepts, and maximize the performance of accelerators based on existing concepts and technologies.  For  Energy Frontier applications this means supporting the development of new techniques that will increase the accelerating gradients so future machines are more compact and less costly. The options considered in our study include acceleration in plasma structures, using either laser or beam driven wakefields, dielectric structures driven by lasers or RF (GHz), the development of new lepton collider designs such as muon colliders and two-beam acceleration, and optimization of existing technologies such as superconducting rf cavities. For  Intensity Frontier, simulations are essential in developing and optimizing integrated designs in order to minimize beam losses due to instabilities caused either by beam self-interactions or by interactions of the beam with the accelerator structures or other media present in the beam pipe.  This  includes both designing mitigation techniques and determining optimal operational parameters.  Hadron colliders at the Energy Frontier have similar requirements, although self-interactions are not important and beam-beam interactions (which are similarly computationally intensive) have to be included.  

Simulations of accelerators for both the Energy and the Intensity frontier are computationally demanding because they often involve a wide range of time and length scales and a wide spectrum of interoperating physics components. For example, high intensity proton drivers of the order of $10^3$ m, operating at an EM wavelength of $10^2$--$10$ m with components of the order of $10$--$1$ m must resolve particle bunches of the order of $10^{-3}$ m. Similarly, laser-plasma accelerators (LPA) of the order of $1$ m in length must resolve laser wavelength and electron bunch size of the order of $1$ $\mu$m.

Most of software for accelerator science are already  highly parallelized and scalable to $> 10$k cores on HPC. They use a wide variety of numerical models, such as electrostatic (multigrid, AMR multigrid, spectral), electromagnetic (finite difference, finite element direct and hybrid, extended stencil finite-difference, AMR finite-difference, spectral), quasi-static (spectral), and  matrix solvers, Particle in Cell, meshin and other libraries, and a variety of analysis tools. In addition, there are ongoing R\&D efforts to port these numerical models on new architectures such as GPU based machines.  Although the physics models implemented in today's simulation tools utilize the above sophisticated infrastructure, because of the size of the computation, often "single physics" or "few physics" models are included in a run. The different physics effects are studied separately, as if they were independent.  This is not the case in general, affecting our ability to find optimal design and operational parameters.  More efforts are needed to integrate multiple physics for more accurate simulations, with the ability to utilize massive computing resources beyond the capabilities of today. In the energy frontier, where single components of the accelerator are simulated separately, end-to-end simulations and integration between components is needed.  For example, plasma based accelerators simulations must be advanced from modeling current experiments at the 10 GeV and 0.1 micron emittance level to future collider concepts involving 100s of stages at the 0.01 micron emittance level, which also requires integration of additional physical models such as scattering and radiation. For high-intensity circular proton machines, a large number of macro-particles ($\sim 10^9$) must be used in the simulations in order to accurately represent \% level losses. In addition, detailed models of important components relevant to all frontier applications are missing from our simulation toolkits because of prohibitive computational cost and complexity (for example target modeling, including Gas dynamics, MHD, and heat loading/dissipation must be integrated to our toolkit). 

\section{Energy Frontier}
\label{sec:acc-sc-ef}
\subsection{Plasma based accelerators  for future colliders}

Laser driven plasma accelerators  (LPAs) \cite{EsareyRMP09} and particle beam driven plasma accelerators (PWFAs) have the potential to reduce the size of future linacs for high energy physics by more than an order of magnitude, due to their high gradient.  High quality GeV beams have been produced by LPAs in 3 cm, while  energy gains of 40 GeV in a meter have been demonstrated in PWFA.   LPA research is in progress at facilities, including the BELLA PetaWatt laser at LBNL, towards high quality 10 GeV beams and staging of multiple modules, as well as control of injection and beam quality. PWFA research includes FACET at SLAC, which is exploring controllable acceleration of high quality electron/positron beams in a meter long PWFA.  These aim to address physics and R\&D challenges for a detailed design of future collider concepts (see white paper by J.P. Delahaye \textit{et al.}).  

Simulations must resolve plasma formation, driver beam propagation and energy transfer, the injection and evolution of high quality particle beams, and the loading of the plasma structure by the beam.  Core methods are explicit and implicit particle in cell and fluid.  These scale well but stretch computational abilities even for current experiments at the 1-10 GeV level in m-scale plasmas with 0.1 micron emittance (including loading of the plasma by the accelerating beam) \cite{web}.  

The path towards high-energy physics applications will likely involve hundreds of 10 GeV-scale stages with injectors, compact beam transport between stages, cooling, and focusing (e.g. adiabatic plasma lens) \cite{Leemans09}.  For collider emittances, this simultaneously increases the length of simulation and the accuracy with which beam emittance must be resolved by one to two orders of magnitude, while domain size increases only modestly.  Also required are simulations, self consistently with the plasma, of scattering, radiation, spin polarization and production of positrons (or other accelerated particles).  These in turn require increased particle number for statistics.  Control of injection or dephasing, or near-hollow channels to mitigate scattering-induced emittance growth, require that plasma formation codes be developed to account for 3D effects and self consistent laser heat deposition.   High average power at kHz-MHz repetition rates will require inclusion of target heat flow and laser modeling.

While scaling can increase particle number and resolution, needs for increased run length and accuracy with added physical models motivate new methods.  Recent examples include computation in a boosted frame, where the scale disparity is reduced, envelope codes which average over the laser period, and methods to reduce unphysical momentum spread.  Numerical methods with improved accuracy and reduced unphysical momentum contributions will be critical.  These may include Vlasov and/or models that exploit specific physics features (e.g. envelope, boost, r-z).  In particular, as compute power appears likely to increase faster than bandwidth, more accurate methods allowing longer timesteps (even if at higher computational cost) may be advantageous. Emerging multicore or SIMD systems function well with PIC codes, but development of common compilers and tools are a high priority for productivity.  Heterogeneous decomposition will also likely be required.

\subsection{Dielectric structures}

Dielectric structures have been found promising in both the GHz
and optical wavelengths, with both types of structures relying on
photonic structure principles of frequency-selective confinement.
In the GHz range, photonic structures formed from arrays of
dielectric rods have been found with high Q values for the
accelerating mode but with reduced wakefields due to the lack of
confinement of higher-order modes.  In the optical wavelengths,
the dielectric breakdown field is in the range of 10 or more
GeV/m, and so hold out the promise of acceleration gradients that
are two orders of magnitude greater than conventional systems.
In the optical, structures vary from 3D, such as the woodpile, to
dielectric fibers, which are cheap, as they are used by the
telecom industry.

With many principles and ideas of using dielectric structures
having been elucidated, there is now a need for assessing many
practical issues.  These include issues of pure electromagnetics,
such as how to efficiently couple energy into these structures
and what structures have sufficient Q values, through
self-consistent effects, such as whether there are instabilities
due to wake fields.  For such studies, algorithmic advances are
needed, with one direction being the need for rapid geometry
layout and meshing algorithms for these complex structures, as
well as fast, scalable, and accurate algorithms able to compute
$>10^9$ degrees of freedom.  As well, algorithms need modification
to take advantage of the many computational accelerators and
advanced instructions (GPU, MIC, AVX2) now or soon available.
Moreover, it is important to develop this software in a
maintainable fashion, which cannot be writing different
implementations for each new architecture.

As well, there is a need for integration of optimization to find
the systems with the best coupling, highest Q, lowest wake
fields, etc.  Such optimizers need to be tailored to the type of
simulations.  For example, optimizers based on differentiation
may not work well with some simulations that have significant
error or particle noise.

As the field progresses, there will be a need for multi-physics.
Because the electromagnetic field deforms the structures, there
is a need for electro-acoustic couplings, and because it heats
the structure, electro-thermal coupling is additionally needed.

With the above developing tool suites, there is a need finally to
carry out the extensive studies of these systems.  There are many
configurations now (RF, optical; 3D woodpiles, gratings; 2D rods,
fibers) with many parameters to vary.  Optimization campaigns are
needed, but they cannot be done blindly.  With so many
parameters, physical intuition will also be important.  Hence,
there will need to be a partnership among computational
physicists, algorithm developers, and computer scientists to
bring the promise of this field to fruition.

\subsection{Muons colliders}

The mission of the Muon Accelerator Program (MAP) \cite{map} is to develop and demonstrate the concepts and critical technologies required to produce, capture, condition, accelerate, and store intense beams of muons for Muon Colliders and Neutrino Factories. The goal of MAP is to deliver results that will permit the high-energy physics community to make an informed choice of the optimal path to a high-energy lepton collider and/or a next-generation neutrino beam facility. Coordination with the parallel Muon Collider Physics and Detector Study and with the International Design Study of a Neutrino Factory will ensure MAP responsiveness to physics requirements.

For a muon colliders an essential computational need is the optimization of cooling channels. Muon cooling is required to reduce the beam phase space so the beam can be efficiently accelerated and so a muon collider will have increase luminosity. A typical muon-cooling channel is 200-300 meters long, and the interaction of the beam with the matter in the absorbers is an essential aspect of its operation. Simulations of such channels typically require about 1-5 CPU-seconds per event, and about a half-million events are required to obtain good statistical accuracy (a substantial fraction of the muons are lost or decay). An optimization run with perhaps a dozen free parameters typically requires several thousand iterations, each of which requires about a million CPU-second, totaling on the order of $10^9$ CPU-seconds. 

Recent parallelization of the simulation codes  have allowed orders of magnitude speed up by running on HPC. In addition to cooling channel target, front end, acceleration, collider, decay rings, and MDI all also require significant modeling, increasing the computing needs required. In turn, muon collider simulations will also require the integration of more physics phenomena such as single particle optics, space charge effects, beam-beam effects and other collective effects. Interaction of the beam with plasma in gas-filled cavities and other materials must also be considered, as well as radiation, particle decay, etc.

\section{Intensity Frontier}
\label{sec:acc-sc-if}

Circular accelerators are a central feature in nearly all
proposed plans for the future of the Intensity Frontier. Since
the intensity-limiting effects in accelerators are collective in
nature, accurate studies of potential collective effects are
critical portions of the design process. The three primary
collective effects in question are space charge, impedance and
electron cloud. Ongoing studies of these effects in the Fermilab
Booster and Main Injector provide concrete examples of the types
of studies that will be necessary for any Intensity Frontier
circular accelerator.  Useful simulations have three main
requirements. The first requirement is a model of the accelerator
itself that contains enough detail to effectively capture the
physical effects leading to losses. Important details include
realistic apertures, magnet fringe fields, misalignments, etc.
The second requirement is a simulated time period long enough to
capture the various loss mechanisms that come into play. The
third is an overall level of fidelity in the simulation great
enough to have confidence in the final results. These three
requirements together put constraints on both software, which
must accommodate the complexity required for realistic models,
and computing hardware, which must be capable of delivering
detailed simulations in a timely manner.  For simulations of the
Main Injector, the first needs are accurate simulations of space
charge and impedance combined with a detailed model of the
accelerator. Space charge and impedance-related simulation topics
to address in the Main injector include space-charge tune shifts
and tune spreads. These studies will lead into studies of the
variation of operational parameters to minimize losses, which
will require many runs as the parameter space is scanned. Further
studies include simulations of injection beam painting and
mitigation techniques such as electron lenses. In all cases, the
simulation program must be benchmarked against corresponding beam
studies (for example at facilities such as ASTA and UMER). 
Simulations of relatively new technologies
such as electron lenses are especially important to pair with
experimental measurements to validate models.

Electron cloud development and its effect on beam dynamics is
another subject of concern in the Main Injector. A program to
simulate electron cloud development is ongoing; simulations of electron
cloud effects in beam dynamics are planned. Because the electron
cloud phenomenon is the product of a complex set of physical
effects, an experimental program to study these effects and
validate the simulations is also required. Such an effort has
started, but will require more work to reach the accuracy needed.
Simulations of space charge and impedance effects in the Booster
are also of great importance. In the Booster, inter-bunch
communication through impedance has been shown to critically
depend on the number of bunches present in the simulation. These
results indicate that simulations containing the entire 84-bunch
filled Booster ring are necessary. Such simulations have all the requirements
of the single-bunch simulations with additional computational
complexity proportional to the number of bunches. Such
simulations require supercomputer resources.

The computational requirements of these simulation programs can
be estimated for the Main Injector. A single Main Injector
revolution takes roughly 11 microseconds. The time scale
associated with losses is half a second. The simulations must
therefore address ~50,000 turns. Simulations must take into
account the variation of the beta function around the ring,
sampling several times per period. There are 104 such periods in
the main injector, meaning that the simulations must contain on
the order of 500 steps per turn. Detailed simulations must
therefore contain tens of millions of time steps. The Intensity
Frontier puts strict limits on acceptable losses. If we take 1e-4
has the acceptable loss limit, and we require 1\% accuracy in our
simulations of loss, we require 1e8 macroparticles. Such
simulations are appropriate for today's supercomputers; when the
additional factors associated with multi-bunch simulations are
added to the mix, simulations will require the very cutting-edge
of current supercomputer technology.

Although we used IF accelerators as an example to
describe the code capabilities and size of computation necessary to move in the
future, similar requirements exist for EF hadron colliders, such as VLHC and 
HL-LHC, although self-fields are not important, and beam-beam effects could be
important.  In addition, for operation of such IF or EF machines of the future
control room feedback capabilities is desirable (because of the loss 
implications).  The necessary analysis workflow
and synthetic diagnostic tools would be similar to those used by HEP
experiments today, since they will have to model the beam detector
response and maintain and correlate the information of the
simulated physics variables to those smeared by the model of the
diagnostics. 

\section{Conventional accelerator technology for both Energy and Intensity Frontier applications}
\label{sec:acc-sc-all}
Conventional accelerator technologies play an important role in
the design of future accelerators both in the Energy Frontier
(EF) and the Intensity Frontier (IF). These technologies, which
have been proven to work in existing accelerators, include the
normal conducting rf and superconducting rf acceleration schemes.
Electromagnetic simulations of accelerator components and systems
are essential to the design and optimization of these machines.
In particular, virtual prototyping of accelerator components and
systems through high performance computing enables accelerator
builders to shorten the time for the design and build cycle,
which will substantially reduce the cost for achieving an
optimized design that satisfies beam quality preservation and
machine operational reliability. These machines include

\begin{itemize}
 \item A high-intensity proton source based on superconducting rf
   technology (IF)
 \item A linear electron-positron collider based on superconducting
   rf technology, capable of delivering 500 GeV -- 1 TeV center of
   mass energy (EF)
 \item A linear electron-positron collider based on high-gradient
   normal conducting rf technology and two beam acceleration techniques,
   capable of delivering 500 GeV -- 3 TeV center of mass energy (EF).
   The computational issues in electromagnetic modeling and simulation
   related to these machines are as follows.
\end{itemize}

For superconducting rf technology that is used in the linacs of
proposed accelerators such as Project X and ILC, the accelerator
cavity is designed to minimize the effects of high-order-modes
(HOMs) to maintain beam stability and to limit extra heat losses
on cryogenics. However, during the fabrication process, the SRF
cavity is deformed from its designed shape because of loose
machining tolerance and the tuning procedure, the HOM properties
such as their external Q can be substantially changed to cause
beam breakup problems at the currents well below the
designed threshold. Furthermore, misalignments of the cavities in
a cryomodule will affect the wakefield even though the
imperfection effects in a single cavity is well understood.
Simulation using the capacity of supercomputers will be an
invaluable tool to study these effects and will provide insights
of how to mitigate any possible problems.  In addition, the
 statistical analysis for a wide range of the scales and
types of deformation and misalignments in these structures
require thousands of computers runs to give a reliable account of
wakefield effects during machine operation.

Another limiting factor that prevents the accelerator from reaching
high gradients is the generation of dark current, which arises
from field emissions of electrons from the surface of an
accelerating structure and their subsequent movement whose
trajectories are determined by the accelerating rf field. Dark
current may lead to beam loading of the accelerator structure
and, if captured, may also produce undesirable backgrounds
downstream in the detector at the interaction point. Therefore,
understanding the mechanism of dark current generation and
capture is essential to the successful development of high
gradient structures for linear colliders such as the CLIC. Also,
it was found experimentally that dark current generation was
enhanced during the transient of the drive power pulse.
Therefore, it is important to perform a time-domain simulation
with a realistic driving pulse to determine the dark current
effects. The capture of dark current downstream may take a long
distance that may involve multiple accelerating modules. The
number of time steps and the number of particles for tracking
needed for these large-scale simulations requires tens of
millions CPU hours on state-of-the-art supercomputers.

In addition to electromagnetic properties, the
studies of thermal and mechanical properties are necessary for
the full design of a cavity. One first calculates the
electromagnetic properties of an accelerating mode in the cavity,
and then uses them to determine the thermal and/or mechanical
properties, which may be used to evaluate the changes in
electromagnetic properties due to thermal expansion or mechanical
deformation. For the integrated simulation of cavities, the study
of these multi-physics effects is a computationally challenging
problem due to the complexity of the cavity geometry and the
different types of physics. The complexity of the modeling arises from 
the fully-dressed SC
cavity together with fast tuner, slow tuner, rf coupler and
helium vessel, as well as the connections to cavity string
installation inside the cryomodule. In addition to improving
existing thermal solvers to handle various boundary conditions
form the external thermal loads, new parallel mechanical solvers
are needed to address important effects such as microphonics. This
new development will provide a transformative tool that can
facilitate a full design optimization of the machine including
all the details and complexities that are involved in the system.

\section{Accelerator modeling science needs}

Thanks to sustained advances in hardware and software
technologies, computer modeling is playing an increasingly
important role for particle accelerators, making it logical to
strengthen programmatic activities. Numerous simulation codes
have been developed, and with a few notable exceptions (e.g., the
SciDAC funded collaboration) the development paradigm has largely
been: a code linked to a project or specialized topic, developed
by a researcher (usually a physicist), with occasional help from
computer scientists.  Maximizing scientific output per dollar
means maximizing the usability of the pool of codes while
minimizing spending on development and support, including through
reductions of duplication/increases in modularity and code
interoperability.

Development and application of accelerator algorithms and codes
have become extremely complex and specialized endeavors, calling
for teams including computational physicists (SciDAC but
expand...), applied mathematicians and computer scientists. Such
an approach is being adopted elsewhere, and calls for a higher
level of coordination among modeling and code development
efforts, with progressive integration of codes into a tool set.
This is all the more timely as computer architectures are
transitioning to new technologies, requiring adaptation.
Separation between software for personal computers versus
supercomputers is also diminishing as the former become multicore
and the later commodity based, and it is essential to envision
tools that function well on a broad range of platforms.

A high-level scripting interface for rapid development and
prototyping, offering easy interfacing with high performance
languages and expandability, represents one solution to the
challenge of coupling of existing codes while minimizing
disruption and enabling both interoperability and expansion
capabilities. With such a construct, existing codes continue
unmodified preserving the very significant investments in
existing accelerator modeling tools, while their functionalities
are exposed to allow combined use for multi-physics simulation.
In the past decade, the Python scripting language has emerged as
a high-level solution for rapid development and prototyping which
can be easily coupled to the high-performance programming
languages.

Current practice is also less than optimal in that with few exceptions
the users of HPC accelerator codes are the developers.  Scientific
productivity would be enhanced by making the accelerator codes more
widely usable.  To do this, techniques include simplified problem setup and submission
through graphical user interfaces with client-server technology, as
has already occured in, e.g., the HPC visualization community.

\section{Summary}
\label{sec:acc-sc-summary}
The computational and computing needs for supporting Accelerator
Science are dominated by the need to optimally utilize
High-Performance-Computing (HPC) and the availability of tools
and resources that makes this utilization possible.  Here HPC is
defined in the conventional way, where parallelism and fast
interconnect is essential to the computation, since each
simulation step requires communication between thousands to
millions of processors. The projected computing  needs for all
the major modeling applications from both energy and intensity
frontiers is shown in Table~\ref{tab:CompNeeds}, where the units
are based on the current performance of our codes on Hopper at
NERSC. Note that in this report we do not detail data storage and
networking needs (with one exception), because our area will not
drive the overall requirements, which are dominated by HEP
experiment needs.  We will leverage from the solutions
implemented to support these programs.

A common theme from the requirements communicated both by our
user community (accelerator scientists operating machines or
performing R\&D at test facilities) and computational accelerator
physicists, is the need for programmatic coordination and support
of code development and computing R\&D to create a sustainable
computational accelerator science program.  Porting of our
algorithms and workflows to new computing architectures
(light-weight CPU plus accelerator) and the R\&D necessary to
create and evaluate new algorithms is an important component of
such coordinated program (including close interactions with HPC
centers to utilize test-beds of new architectures). An example of
such programmatic support today is the SciDAC program, although
it is desirable that in the future there is more focus on the
specific physics solutions needed to further develop our tools.  Another
common theme is the need for supporting the development of
community libraries and tools, including standardized user
interfaces, geometry and data descriptions, I/O and analysis tools.
Because our applications require true HPC capabilities,
development of generic workflow tools that perform in an HPC
environment as well as local workstations and clusters is very important, as is the development and
integration to our toolkit of parameter optimization libraries,
that will be available across all HPC platforms.  The development
of such environment will enable experimentalists and machine
operators to take advantage of these computational capabilities
and will be essential in training students and young researchers
to help develop the new accelerator concepts and technologies
that will move the field of particle accelerators forward.   In
addition, it is essential for such a program to support and
coordinate physics model validation and verification, ultimately
with comparisons to experimental data of well controlled
experiments in test facilities or operating accelerators.

Intensity Frontier machines of the future require control room
feedback capabilities (because of the loss implications), a
capability that is also important to Energy Frontier test
facilities (for guiding and interpreting experiments).   Would it
be possible with utilization of new computing technologies to
deliver such fast turnaround?  The challenge on both the
performance of the computational tools and the availability of
computing resources becomes even  more daunting if we consider
the need to analyze the simulated data in order to extract useful
information.  The analysis of the simulated data ($\sim$ TB) has
to produce the same quantities observed by the beam diagnostic
detectors.  Note that this is a more general requirement, because
it is necessary for accurate comparisons of simulated and
observed data independently of the ability to do that in ``almost
real-time'' in the control room.  The necessary analysis workflow
and synthetic diagnostic tools similar to those used by HEP
experiments has to be developed to properly model the detector
response and maintain and correlate the information of the
simulated physics variables to those smeared by the model of the
diagnostics.  Such analysis tools have to be HPC capable, to
allow for the fast turnaround necessary for control room
feedback, and they will also require development of new models
and algorithms.  Finally, this is probably the only application
in accelerator modeling that data transfer speed and data
availability, storage, and cataloging has similar requirements to
those of a HEP experiment DAQ system.

Although different applications have different specific
requirements for the development of new or more efficient physics
or computational models, all of them require integrated
multi-scale, multi-physics modeling.  Currently, for
high-fidelity modeling, because of the many degrees of freedom
involved in the problem, we run "single physics" or "few physics"
model simulations.  Such simulation environment require the
development of efficient numerical algorithms and models able to
utilize massive computing resources and the availability of such
resources at the HPC centers. Deployment of such capabilities
will enable end-to-end simulations to validate designs based on
new concepts and end-to-end operational parameter optimization of
accelerators about to be commissioned.  It should be noted that
in some cases end-to-end modeling involves integration of physics
and numerical models developed for different applications (for
example, for a plasma based accelerator consisting of many plasma
stages, both plasma physics tools and conventional beam-dynamics
tools have to be used in the model to produce an optimal
solution).

Intensity Frontier accelerator needs are dominated by the need to
control and mitigate beam losses.  This demands both careful
design of the accelerator structures and accurate modeling of
beam-halo (and its creation mechanisms) and the accelerator
geometry (apertures) and accelerator elements fields and
positions.  This implies tracking many bunches of $\sim 10^9$
macroparticles per bunch for $\sim 10^5$ turns including
self-fields, impedance effects, and bunch-to-bunch interactions.
Finding the optimal parameters of operation will require
end-to-end  optimization runs, while developing mitigation
techniques possibly requires the implementation of new physics in the HPC
environment, to model the new components (for example, electron
lenses for space-charge compensation). Energy Frontier
application based on protons have similar modeling needs for loss
control and mitigation, although in this case impedance effects
dominate (and possibly beam-beam interactions in a collider) as
self-interactions are not important.

Energy Frontier accelerator needs are dominated by the need to
develop end-to-end simulations to characterize and optimize beam
stability and emittance and transport efficiency.  New
accelerator concepts have many specific new physics model
capability needs, for example development of electromagnetic
plasma and beam methods capable of resolving 0.1 km-scale
propagation of 10 nm scale emittance bunches and laser drivers,
and the corresponding bunch conditioning and focusing, but there
are also many common needs.  For example, radiation and
scattering are relevant to muon collider, plasma and
gamma-gamma options, and modeling of targets,  including Gas
dynamics, MHD, and heat loading/dissipation are relevant
to both EF and ID applications.  Developing these new models
demands R\&D both on the physics and numerical algorithm area.
Because of the physics requirements imposed by some of the new
concepts considered, minimization of numerical noise is very
important in these applications.  This constraint has a direct
impact on the choices of numerical techniques for different
physics implementations.  Plasma accelerators additionally require 
computation of these effects with accurate plasma and laser dynamics,
 often requiring unique algorithms.

\begin{table}[t]
\begin{center}
\begin{tabular}{|l|l|} 
 \hline 
 Computation (Mhours) & $15000$ \\ \hline
 Typical cores for production runs & $50000$ \\ \hline
 Maximum cores for production runs & $5$M\\ \hline
 Data read and written per run (TB) & $1000$\\ \hline
 Minimum I/O bandwidth & $100$ GB/sec\\ \hline
 Memory requirement per core & $0.2$ GB \\ \hline
Shared file-system space (on site) & $6$ PB\\ \hline
Shared file-system space (distributed, cataloged) & $60$ PB\\ \hline
\end{tabular}
\caption{Compute needs in 10 years.}
\label{tab:CompNeeds}
\end{center}
\end{table}

\end{document}